# From Chalkboards to Chatbots: SELAR Assists Teachers in Embracing AI in the Curriculum

**S**: Supportive **E**: Education **L**: Lectures **A**: AI **R**: Resource

Hani Alers [a], Aleksandra Malinowska [b], Mathis Mourey [a,c], Jasper Waaijer [a]

[a] The Hague University of Applied Sciences, The Hague, The Netherlands

[b] University of California, Santa Barbara, CA, USA

[c] Univ. Grenoble Alpes, Grenoble INP, CERAG, 38000 Grenoble, France

## Abstract

This paper introduces SELAR, a framework designed to effectively help teachers integrate artificial intelligence (AI) into their curriculum. The framework was designed by running workshops organized to gather lecturers' feedback. In this paper, we assess the effectiveness of the framework through additional workshops organized with lecturers from the Hague University of Applied Sciences. The workshops tested the application of the framework to adapt existing courses to leverage generative AI technology. Each participant was tasked to apply SELAR to one of their learning goals in order to evaluate AI integration potential and, if successful, to update the teaching methods accordingly. Findings show that teachers were able to effectively use the SELAR to integrate generative AI into their courses. Future work will focus on providing additional guidance and examples to use the framework more effectively.

## 1. Introduction

Artificial intelligence (AI) is increasingly recognized as a transformative force in education, offering unprecedented opportunities to enhance teaching and learning processes (Luckin, Holmes, Griffiths, & Forcier, 2016; Zawacki-Richter, Marín, Bond, & Gouverneur, 2019). AI technologies have the potential to personalize learning experiences, automate administrative tasks, and provide intelligent tutoring systems that adapt to individual student needs (Chen, Chen, & Lin, 2020; Holmes, Bialik, & Fadel, 2019). These advancements can lead to improved educational outcomes by catering to diverse learning styles and enabling educators to focus on higher-order teaching activities (Tuomi, 2019; Miao, Holmes, Huang, & Zhang, 2021).

Despite the promising benefits, integrating AI into educational settings presents significant challenges. A critical barrier is the lack of preparedness among educators to effectively incorporate AI technologies into their curricula (Celik, Dindar, Muukkonen, & Järvelä, 2022; Ayanwale, Adelana, Molefi, Adeeko, & Ishola, 2024). Many teachers report insufficient knowledge and skills related to AI, leading to uncertainty and resistance toward adopting these tools (Williamson, 2016; King & ChatGPT, 2023). Moreover, concerns about academic integrity have emerged with the advent of generative AI tools capable of producing human-like text, raising questions about assessment and plagiarism (Cotton, Cotton, & Shipway, 2024).

Existing literature emphasizes the necessity for targeted professional development and support frameworks to empower teachers in embracing AI technologies (Lin & Van Brummelen, 2021; Vazhayil, Shetty, Bhavani, & Akshay, 2019). However, there is a paucity of practical tools that guide educators in updating their courses to reflect current AI advancements while maintaining educational rigor and integrity (Zawacki-Richter et al., 2019). Addressing this gap is essential to ensure that students are equipped with relevant skills and that educators can confidently integrate AI into their teaching practices (Holmes et al., 2019).

In response to this need, the present study introduces the SELAR framework—a structured tool designed to assist educators in systematically evaluating and updating their learning goals and teaching materials in the context of modern AI developments. The SELAR framework operates in two key stages: first, determining whether existing learning goals are robust against AI solutions; and second, deciding how AI tools can be effectively integrated into the curriculum to enhance learning outcomes.

In addition, this study evaluates the effectiveness of the SELAR framework through structured workshops with lecturers from The Hague University of Applied Sciences (THUAS). The participating lecturers were introduced to the framework and guided in applying it to their own courses. Subsequently, they deployed the updated courses to teach students in the university and provided feedback on their experiences through interviews. By doing so, the study evaluates the impact of SELAR in the field and aims to contribute a practical solution to the challenges faced by educators in integrating AI into their curricula.

The remainder of this paper is organized as follows: Section 2 presents the SELAR framework and elaborates on its design. Section 3 outlines the methodology used in evaluating the SELAR framework's effectiveness. Section 4 presents the results of the evaluation, followed by a discussion in Section 5. Finally, Section 6 concludes the paper and suggests directions for future research.

## 2. Explaining the SELAR framework

The SELAR framework is an innovative tool designed to assist educators in adapting their curriculum to ensure it copes with recent generative AI developments. As Celik et al. (2022) note, AI's potential to improve teaching practices and educational outcomes is contingent upon addressing challenges such as AI-focused teacher training, ethical considerations, and integration into existing educational frameworks. Continuous research and robust support systems are crucial for the successful implementation of AI in education. Furthermore, McArthur et al. (2005) point out that involving teachers in the curriculum development process is essential for ensuring that AI educational tools are both relevant and accessible. This alignment with the needs of teachers facilitates the successful integration of AI in classrooms. We designed the SELAR framework to be teacher-focused for that very reason.

This framework empowers teachers to integrate AI concepts into their courses, thus preparing students for a technology-driven world. By systematically evaluating and updating the learning goals of the courses, the framework ensures that educational content remains relevant and effective in the context of modern AI advancements. The process involves two key workflows when updating a given course.

## 2.1. Workflow Learning Goals: updating learning goals

The first workflow (shown in Figure 1) reassesses each learning goal of the course to identify potential areas for AI integration and ensure assessments remain AI-proof. The workflow outlines the process of AI-proofing learning goals and utilizing AI tools in an educational setting. The workflow can be applied to any learning goal in a given course. The workflow has 15 blocks as explained below[1].

### 1. Examine the learning goal

Start by conducting a detailed evaluation of the learning goal. Consider its relevance in today's educational environment, where students have access to advanced AI tools. Determine whether the goal addresses essential skills and if it aligns with the current learning outcomes. This step is crucial to ensure that the goal remains aligned with the professional competencies required in the field, particularly in areas where AI may already play a role.

### 2. Can part of the learning goal be automated with AI?

In this block, evaluate if certain tasks within the learning goal can be automated through AI tools. For example, tasks such as generating reports, running simple statistical analysis, or summarizing content may be handled by AI. However, you must carefully assess whether automating these tasks will still allow students to grasp the essential competencies the course aims to teach for the rest of their education.

- If Yes, move to Block 3 to consider whether AI usage aligns with the learning goals.
- If No, proceed directly to Block 11 to finalize the learning goal without AI intervention, ensuring that the goal remains hands-on and relevant to student learning.

### 3. Is it acceptable to use AI assistance for this learning goal?

This block involves a deeper analysis of the educational value of allowing AI to perform certain tasks. Consider whether all the skills covered by this learning goal are crucial for students to master manually or if some can be supported by AI. You need to reflect on whether reliance on AI would detract from the core learning outcomes or complement them. This decision requires an understanding of whether automating the task fits into the broader educational context.

- If Yes, proceed to Block 6 to adapt the learning goal to incorporate AI.
- If No, move to Block 4, where you will explore how to prevent students from using AI to bypass important aspects of the learning goal. Ensuring that students are learning the intended skills is crucial in this case.

### 4. Is AI-proof assessment possible?

---
[1] Use Figure 1 provided in the Appendix to follow through the process of the workflow.

In this block, determine if you can design assessments that cannot be easily completed by AI. These assessments should test the students' comprehension and skills without the risk of AI-generated answers. Personalized assignments, oral exams, and in-class evaluations are examples of methods that can ensure students demonstrate their own understanding. It is essential that students cannot use AI tools to complete critical tasks that they need to master themselves.

- If Yes, proceed to Block 5 to finalize the assessment design.
- If No, move to Block 14, where assistance from the AI Helpdesk may be required. The helpdesk can offer solutions or suggest alternative methods to ensure that assessments maintain their integrity.

5. Make the assessment AI-proof

Design and implement assessments that are resistant to AI automation. This involves crafting assessments in such a way that students must engage directly with the material. The goal is to ensure that AI cannot substitute for student performance, particularly in areas requiring critical thinking, problem-solving, or creativity. By doing this, you maintain the rigor of the course and ensure that learning outcomes are met. AI-proof assessment examples are provided as input in Block 13.

Once the assessment is AI-proof, move to Block 11, as the learning goal is now ready to be finalized and incorporated into the course.

6. Change the learning goal to utilize AI

If AI can be integrated into the learning goal and it is educationally appropriate, modify the goal to explicitly involve AI tools. This could mean teaching students how to use AI in tasks they will encounter in the workplace, such as automating reports or conducting research. Incorporating AI helps students build relevant skills for future careers while also enabling instructors to focus on more advanced aspects of the subject.

- After adapting the learning goal to utilize AI, move to Block 7 to see if using AI has changed the workload of the student. If students have less work after utilizing AI then the level of some learning goals can be increased, ensuring the overall course remains challenging.

7. Can the level of another learning goal be increased?

Assess whether, by automating certain tasks with AI, the difficulty or depth of other learning goals in the course can be increased. AI assistance may free up instructional time, allowing you to demand higher levels of proficiency in critical areas such as analysis, synthesis, or application of knowledge. This ensures that students continue to be challenged and develop key skills that AI cannot easily replicate.

- If Yes, proceed to Block 8 to modify that goal.
- If No, move to Block 9 to explore the possibility of adding a new learning goal that leverages the time saved by AI.

### 8. Increase the level of another learning goal

Adjust another learning goal by raising its difficulty or complexity. Since AI may handle simpler tasks, students can focus on more advanced elements, or completing the same learning goal but to a higher proficiency level. This ensures that AI is used as a tool to enhance learning without reducing the overall rigor of the course.

- Once the level of the learning goal is increased, move to Block 10 to update the course materials accordingly.

### 9. Add a new learning goal

If AI automation has freed up time or resources, this is an opportunity to introduce a new learning goal. This could focus on areas such as AI literacy, ethical use of AI, or new skills that were previously unaddressed. Adding new goals keeps the course relevant and ensures that students are equipped with the necessary skills for the evolving landscape of their field.

- Once the new goal is added, proceed to Block 10 to update the learning materials to reflect the changes.

### 10. Use the next workflow to update learning materials

After adjusting the learning goals, review and update the course materials to align with these new goals. This could include new readings, assignments, and resources that support the updated learning outcomes and integrate AI where appropriate. Ensuring that the course materials reflect the latest changes guarantees that students have the necessary resources to engage with the modified curriculum.

- Once the materials are updated, move to Block 11 to indicate that the process for this learning goal is complete.

### 11. Done

After reviewing and updating the learning goal, mark it as complete. At this point, you can move on to the next learning goal and repeat the process from the top. This ensures that all learning goals in the course are systematically reviewed and updated for the current AI-empowered educational environment.

### 12. AI skills checklist

This block serves as a reference provided by the institution, outlining the specific AI skills students are expected to learn or use. Examples include AI capabilities such as report generation, data analysis, or literature review. The checklist helps educators determine which tasks can be automated and ensures that students are proficient in using

AI for appropriate tasks. This checklist is used as input to update the learning goals accordingly.

### 13. AI-proof assessments

This block offers guidelines or tutorials on how to design assessments that are resilient to AI use. These assessments focus on tasks that require human interaction, original thinking, or context-specific solutions that AI cannot easily replicate. It helps course managers design robust evaluation methods that prevent AI from undermining academic integrity.

### 14. Contact AI Helpdesk

Not all staff members are expected to be proficient with the use of AI. If at any point in the process the instructor encounters difficulties, they can reach out to an organization-wide designated AI Helpdesk for support. Such a helpdesk should provide expertise in generative AI tools and offer practical solutions for integrating or mitigating AI use within the course. It acts as a vital resource when standard solutions are not enough to address AI-related issues in the curriculum.

### 15. AI incorporation training

To successfully adapt learning goals and assessments in an AI-driven environment, instructors may need specific training on how to integrate AI tools effectively into their teaching. This training ensures that instructors understand the potential uses and limitations of AI and can implement it thoughtfully within the course design. This is essential for maintaining educational quality while embracing technological advancements.

This workflow should be completed for all learning goals linked to an educational course. If additional learning goals are created, they should also go through the workflow to examine how they should be handled with regards to AI. Once all learning goals are handled, the next phase involves either increasing the difficulty or introducing a new learning goal to maintain the course's challenge level. Following this adjustment, the user proceeds to the next workflow discussed in the coming section.

## 2.2. Workflow Learning Materials: updating learning material

The second workflow (shown in Figure 2) helps in breaking down the updated learning goals from Workflow Learning Goals into specific tasks that are partially automated. The workflow consists of Blocks 16-26 as explained below. Note that Blocks 12, 14, and 15 here have the same description as those blocks with the same numbers in Workflow Learning Goals. Therefore, they will not be explained again here.

16. Skill you want to teach

Each learning goal can be divided into a set of skills or tasks that the student needs to learn in order to master it. Begin by identifying the skill or task that you want students to learn. This is a skill that students need for future work environments. It is important to define the skill clearly before proceeding to determine whether it can be automated.

- Once the skill is identified, move to Block 17 to assess whether this skill can be partially or fully automated using AI.

17. Do you want this skill to be (partially) automated using AI?

Here, you evaluate if the skill can be automated by AI, meaning the student can complete the task using AI tools without necessarily mastering the manual process. This does not mean the student learns the skill by using AI but that AI performs the task itself. The decision here depends on whether learning how to complete such tasks manually is crucial for the student's development or whether it can be left to be handled by automation.

- If Yes, proceed to Block 18 to experiment with AI automation.
- If No, move directly to Block 26 to finish this task without automation.

18. Try using AI to complete a new example of the task

Experiment with using AI to automate the task. Run multiple examples through the AI tool to see how it handles the variations. You may need to return to this step multiple times (as part of the workflow) with different examples to explore how the AI will behave in each case. This experimentation helps determine the reliability and efficiency of using AI for this specific task and helps refine the automation process.

- Once you have tried one example, move to Block 19 to assess the success of the AI automation.

19. Did you (partially) succeed?

This block is at the core of the workflow, as it determines the effectiveness of using AI for task completion. Assess whether the AI managed to complete the task fully or partially. Success here indicates whether the AI can be useful for this skill or if adjustments need to be made.

- If Yes (if you succeeded), proceed to Block 20 to document your approach.
- If No (if you failed), move to Block 21 to note down what went wrong.

20. Note down (or revise) steps to complete tasks

If the task was successfully or partially completed using AI, carefully document the steps you followed. These notes will serve as instructions or a guide for students to replicate the task using AI tools. This step ensures that students can follow a structured approach to using AI for similar tasks.

- After documenting the steps, proceed to Block 22 to detail how you evaluated the AI's output.

21. Note down potential dead end

If you were unable to complete the task, document the reasons behind the failure. Understanding why the AI failed is crucial to preventing students from making the same mistakes. This information also helps guide future attempts at task automation or the decision to avoid AI for this particular task.

- Once the dead end is noted, move to Block 22 to describe how you evaluated the AI's output, even if it was unsuccessful.

22. Document (or revise) how you evaluated output

Here, you explain how the AI's output was assessed. It is essential to teach students how to critically evaluate AI-generated results to ensure they meet the required standards. This step is crucial for students to understand the limitations of AI tools and how to ensure the quality and accuracy of their outputs.

- After documenting the evaluation method, move to Block 23 to review the success or failure of multiple attempts.

23. Did you fail 5 times in a row?

If you repeatedly fail to complete the task using AI, it is likely you have encountered a significant challenge. If this is the case, it may be necessary to seek expert advice from the AI Helpdesk to troubleshoot the issue and find alternative solutions.

- If Yes (if you failed five times in a row), go to Block 14 to contact the AI Helpdesk for support.
- If No (if you have not failed five times in a row), proceed to Block 24 to evaluate your overall success.

24. Did you succeed 5 times in a row?

If you managed to successfully complete the task five times using AI, it means you have likely mastered the process. At this point, you have the knowledge and experience needed to create learning materials that teach students how to use AI for this task.

- If Yes, move to Block 25 to document the methodology for completing the task with AI.
- If No (if you have not succeeded five times in a row yet), go back to Block 18 to keep refining your approach.

### 25. Document methodology to complete task with AI

Compile all your successes and failures into clear instructional material. This will serve as a comprehensive guide for students, including the steps for using AI to complete the task and how to evaluate AI-generated output. This material helps students understand the nuances of task automation and ensures they are prepared to use AI responsibly and effectively.

- Once the methodology is documented, proceed to Block 26 to finalize this task.

### 26. Done

Once you have completed all the steps for this task, mark it as complete. You can now move on to the next task associated with the learning goal and repeat the workflow from the beginning. When all tasks related to the learning goal are complete, return to Workflow Learning Goals to finalize the overall learning goal design.

This structured workflow ensures that each task within a learning goal is thoroughly evaluated for its suitability for AI automation, and if automation is possible, that it is done thoughtfully and effectively. The workflow also focuses on helping students learn how to evaluate AI outputs and avoid common pitfalls.

# 3. Evaluation methodology

To study how lecturers from different academic fields can use the SELAR framework to integrate AI in their teaching practice, we designed structured workshops. In order to keep up with the rapid pace of AI development, such workshops will be conducted on a yearly basis to iteratively improve the SELAR framework. Results presented in this article correspond to the first iteration of improvement.

## 3.1 Workshop design

We used a qualitative research design, focusing on observational methods and participant feedback. More specifically, we run interactive workshops, lasting approximately 1.5 hours, followed by feedback sessions of approximately 0,5 hours, allowing for in-depth exploration of lecturers' experiences with the framework. The feedback sessions are a combination of structured questionnaire and semi-structured interviews.

The workshops are held in person at a THUAS Applied Sciences campus to facilitate accessibility. Each session lasts approximately two hours and is structured as follows:

1. **Pre-workshop preparation**: Lecturers receive guidelines to prepare at least one learning goal from their module to apply to the framework.

2. **Initial exploration**: Participants independently explore the framework to build a foundational understanding. Student observers are present to answer questions about the framework and clarify potential wording issues.

3. **Main workshop**: The main workshop consists of two stages:
    - **Stage one**: Lecturers evaluate their prepared learning goals to identify opportunities for AI integration and discuss potential strategies to make assessments resilient against AI misuse, such as through oral or written exams less susceptible to AI interference.

    - **Stage two**: Participants deconstruct their learning goals into specific tasks that could be automated or enhanced using AI tools. This stage encourages practical application and critical thinking about AI's role in education.

4. **Support and observation**: Student researchers take detailed notes on interactions and engagement. Student observers will act as a support person to assist with questions or technical issues.

5. **Feedback session**: A structured questionnaire is first administered to gather feedback on the framework's usability, relevance, and potential improvements. The questionnaire is followed by recorded semi-structured interviews.

## 3.2 Data collection

The sampling procedure is conducted through targeted emails to various departments of The Hague University of Applied Sciences. The emails included all necessary information regarding the purpose of the study and the workshops, encouraging voluntary participation. As it is the first iteration of improvements, we organized the workshops for 5 lecturers from the Mechatronics, Information and Communication Technology (ICT), and International Business departments. The lecturers were of various ages, genders and academic experience. Participants for the following iterations will be recruited through snowball sampling.

We collect data continuously through the workshops. More precisely, the workshops are recorded with microphones, researchers keep notes of what is happening during the workshops and lecturers' feedback was collected through the questionnaire and the interviews.

## 3.3 Data analysis

Audio recordings from the workshops are transcribed verbatim to ensure an accurate and comprehensive capture of the discussions. These transcripts, along with the observational notes and participants' feedback from the questionnaires, are organized and imported into qualitative data analysis software to facilitate efficient coding and theme development.

An open coding process was employed, where the researchers independently reviewed the data to identify significant statements, concepts, and patterns emerging from the participants' interactions and reflections. Codes were then grouped into categories based on conceptual similarities, leading to the identification of overarching themes that represent the collective experiences and perceptions of the lecturers.

Thematic analysis allowed for the exploration of these themes in relation to the study's objectives, revealing how lecturers from various disciplines and experience levels perceive the usability, relevance, and potential impact of integrating AI into their teaching practices. To enhance the credibility and reliability of the findings, triangulation was utilized by cross-verifying data from multiple sources, including transcripts, observational notes, and questionnaire responses. Any discrepancies in coding are resolved through collaborative discussions among the research team, ensuring a consensus on the interpretation of the data.

## 4. Results and analysis

A thematic analysis was conducted to examine the workshops with teachers from various faculties about their opinions on the framework. Initially, we performed open coding on the transcripts to identify meaningful units of data related to their perspectives, challenges, and recommendations. Through iterative analysis, we identified recurring patterns and similarities among the codes. For example, codes such as "positive feedback" and "guidance needed" emerged frequently across multiple workshops.

Themes were developed by clustering related codes into two major categories: positive thoughts about the framework and areas for improvement. The second theme was further divided into two sub-themes: proposed modifications and aspects that were difficult to follow or understand. These themes were refined through constant comparison and discussion among the research team, leading to a comprehensive understanding of the teachers' experiences. The main themes and sub-themes are elaborated upon in the following sections.

### 4.1 Positive thoughts about the framework

The workshop outcomes indicated a universally positive perception of the framework among all participating teachers. They appreciated the structured approach it provided, finding it particularly valuable for integrating AI into their classrooms—an endeavor they had considered but lacked the time or expertise to implement.

The framework was praised as a practical guideline and an effective starting point for adapting existing courses to incorporate AI developments. One teacher expressed satisfaction with its utility, stating, "It's nice to have guidelines to make the course AI-proof, something I had been considering myself." Another remarked, "It perfectly complements existing frameworks for making a learning goal." These comments illustrate the teachers' proactive attitudes and their positive evaluations of the framework's impact on their teaching practices.

### 4.2 Areas for improvements

Despite the overall positive reception of the framework, teachers identified several areas needing improvement to enhance its usability and effectiveness. A prominent theme was the **need for easy-to-follow instructions**. Teachers expressed a desire for more comprehensive guidelines to facilitate understanding and application of the framework. They suggested that supplementary materials, such as informative videos and practical examples, would significantly aid in this process. One teacher remarked, "A video could help. The chart is clear, but it's also that I now know how it works." Another added, "It helps a lot to have some examples; then it becomes much more tangible to use it." These insights highlight the importance of providing clear instructional support to enable teachers to utilize the framework confidently.

The second major theme centered on **recommendations for modifications to the framework** to improve clarity and engagement. Teachers advocated for the inclusion of structural templates to assist with documentation and comprehension. Concerns were raised about the brevity and vagueness of certain instructions, which made initial understanding challenging. As one teacher pointed out, "Some instructions are a bit brief. When you read it for the first time without an explanation, it's difficult to understand. For example, Block 25 of Workflow Learning Materials document is a very broad instruction." Additionally, there was a strong preference for incorporating collaborative elements into the framework. Teachers believed that engaging in group discussions would enhance interactivity and make the learning process more enjoyable. One teacher emphasized this by saying, "I think there need to be a lot of activities where you just sit together at a table and just do it. If you send people home—teachers especially—and say, 'You do this and give it a try and report back in two weeks,' it's not very interactive. You want to have people sit at a table, sit together, discuss matters, and that will make it more lively, more fun as well."

Teachers also identified specific **aspects of the framework that were difficult to understand or follow**. A notable challenge involved the assessment of learning goals, particularly in understanding how to "increase the level of learning goals." The confusion often stemmed from ambiguous terms like "other," which complicated the progression of different learning goals. One teacher noted, "The word 'other' in the block 'increase the level of other learning goal' (Block 8 in Figure 1)—because the teacher would do other learning goals later," indicating uncertainty about how learning goals should evolve over time. Another area of difficulty was related to documenting processes and outcomes. Teachers found the guidelines for effective documentation to be vague, leading to potential inconsistencies in how the framework was applied. A teacher suggested, "You could use a template for it. Attach an extra block with a template on how to document everything," highlighting the need for more structured support in this area.

Further challenges were encountered in developing **AI-proof assessments**. Teachers felt that the existing list of AI-proof assessment methods was incomplete and that additional resources provided by the institution were necessary to select appropriate and effective techniques. One teacher expressed, "Selecting correct ways of assessment—I think that is the most complicated thing," while another admitted, "I need more time for the assessment part." Concerns were also raised about specific components of the framework, such as the "Did you fail five times in a row?" block. Teachers questioned the rationale behind requiring multiple failed attempts and the lack of verification mechanisms. Comments included, "Did you fail five times in a row? So this is where people get very frustrated. If it's a dead end, you're still trying," and "We don't know if five times is enough, and we rather say we accept this margin of error."

In summary, while teachers valued the framework's potential to enhance their teaching practices, they emphasized the necessity for clearer instructions, collaborative opportunities, and specific modifications to address challenges in understanding and implementation—particularly concerning assessment strategies and the framework's more ambiguous elements.

## 5. Discussion

This study explored the effectiveness of the SELAR framework in assisting educators to integrate AI into their curricula. The findings reveal both positive receptions and areas needing improvement, which align with and contribute to the existing body of research on AI integration in education.

### 5.1 Link to literature

The positive reception of the SELAR framework by participating teachers resonates with previous studies emphasizing the need for practical tools to facilitate AI integration in education (Holmes et al., 2019; Zawacki-Richter et al., 2019). Teachers appreciated the structured approach of SELAR, which supports the notion that educators benefit from clear guidelines when adopting new technologies (Lin & Van Brummelen, 2021). This finding aligns with Celik et al. (2022), who highlighted that teacher training and support are critical for successful AI implementation.

However, the participants also expressed a need for additional guidance and resources, such as instructional videos and examples. This echoes the challenges identified by Ayanwale et al. (2024), where pre-service teachers required more comprehensive training to develop AI literacy. The concerns about assessment integrity in the era of AI, particularly regarding generative AI tools like ChatGPT, mirror the issues raised by Cotton et al. (2024) and King and ChatGPT (2023) about academic integrity and plagiarism.

Moreover, the participants' desire for collaborative elements and peer discussions during the implementation process supports the findings of Vazhayil et al. (2019), who advocated for collaborative teacher education approaches to introduce AI in schools. The need for iterative refinement of the framework is consistent with the recommendations of iterative design processes in educational technology (Luckin & Holmes, 2016).

### 5.2 Practical implications

The SELAR framework has the potential to be a valuable tool for educators across various educational contexts. Its structured approach enables teachers to systematically evaluate and update their learning goals and teaching materials in response to AI advancements. By incorporating AI into curricula, educators can enhance learning outcomes, promote AI literacy among students, and better prepare them for a technology-driven workforce.

Implementing the SELAR framework can lead to several practical benefits:

- **Enhanced curriculum relevance**: By regularly updating learning goals to reflect AI developments, curricula remain current and relevant, aligning with industry demands and technological advancements.

- **Improved teaching strategies**: Educators can leverage AI tools to automate routine tasks, allowing them to focus on higher-order teaching activities and personalized instruction (Tuomi, 2019).
- **Student skill development**: Integrating AI into learning goals fosters students' critical thinking and problem-solving skills, particularly in evaluating AI outputs and understanding AI's limitations.
- **Assessment integrity**: The framework encourages the development of AI-proof assessments, addressing concerns about plagiarism and ensuring that evaluations accurately reflect student learning.

For successful implementation, institutions should provide comprehensive support, including training programs, resources, and opportunities for collaboration among educators. Tailoring the framework to specific educational settings and disciplines may enhance its applicability and effectiveness.

## 5.3 Limitations

While the study provides valuable insights, several limitations must be acknowledged:

- **Sample size and diversity**: The study involved only five lecturers from three departments within a single institution. This small and localized sample limits the generalizability of the findings. Future research should include a larger and more diverse group of participants from different institutions and disciplines to enhance the validity of the results.
- **Short-term evaluation**: The assessment of the framework's effectiveness was based on initial workshops and immediate feedback. Longitudinal studies are needed to evaluate the sustained impact of the SELAR framework on teaching practices and student learning outcomes over time.
- **Limited scope of AI tools**: The study focused on generative AI technologies, but AI encompasses a wide range of tools and applications. Expanding the framework to address other AI technologies could provide a more comprehensive approach to AI integration.
- **Potential bias:** Participants may have had varying levels of interest and familiarity with AI, which could influence their perceptions of the framework. Future studies should control for such variables or include measures to assess participants' baseline competencies.

By addressing these limitations, the SELAR framework can be further developed to become a robust tool that significantly contributes to the integration of AI in education.

## 6. Conclusion

AI's potential to improve teaching practices and educational outcomes is contingent upon addressing challenges such as AI-focused teacher training, ethical considerations, and integration into existing educational frameworks. Continuous research and robust support systems are crucial for the successful implementation of AI in education (Celik et al., 2022).

In this study, we presented the SELAR framework, designed to assist teachers in updating their courses to align with advancements in AI. We assess its effectiveness with data collected from five workshops conducted with lecturers from various fields at The Hague University of Applied Sciences. They unanimously appreciated the framework for its structured approach, which facilitates the integration of AI in their teaching. However, they suggest that additional detailed guidelines, including instructional videos and examples, would improve their understanding and application of the framework.

Our research also revealed that the workflows could benefit from some modifications. Lecturers expressed confusion about the assessment of learning goals, specifically highlighting the need for clearer definitions. Additionally, they sometimes found the sequence of steps in Workflow Learning Materials unclear, particularly regarding the order in which multiple prompts should be addressed. Our results align with the recommendation of Lin, P., & Van Brummelen (2021) that prioritizing teacher education and providing comprehensive training programs are essential to equip teachers with the necessary skills and confidence to effectively use AI technologies, ultimately benefiting students.

These results highlight both the strengths and areas for improvement in the framework, underscoring the ongoing journey to optimize AI integration in educational settings. The SELAR framework needs now to be updated using the results of this study and to be tested with a new iteration of workshops with lecturers[2].

---

[2] Find the next iterations on : https://datainnovationhub.eu/projects/selar-workflow/

# Appendix

*Figure 1: SELAR – Workflow Learning Goals*

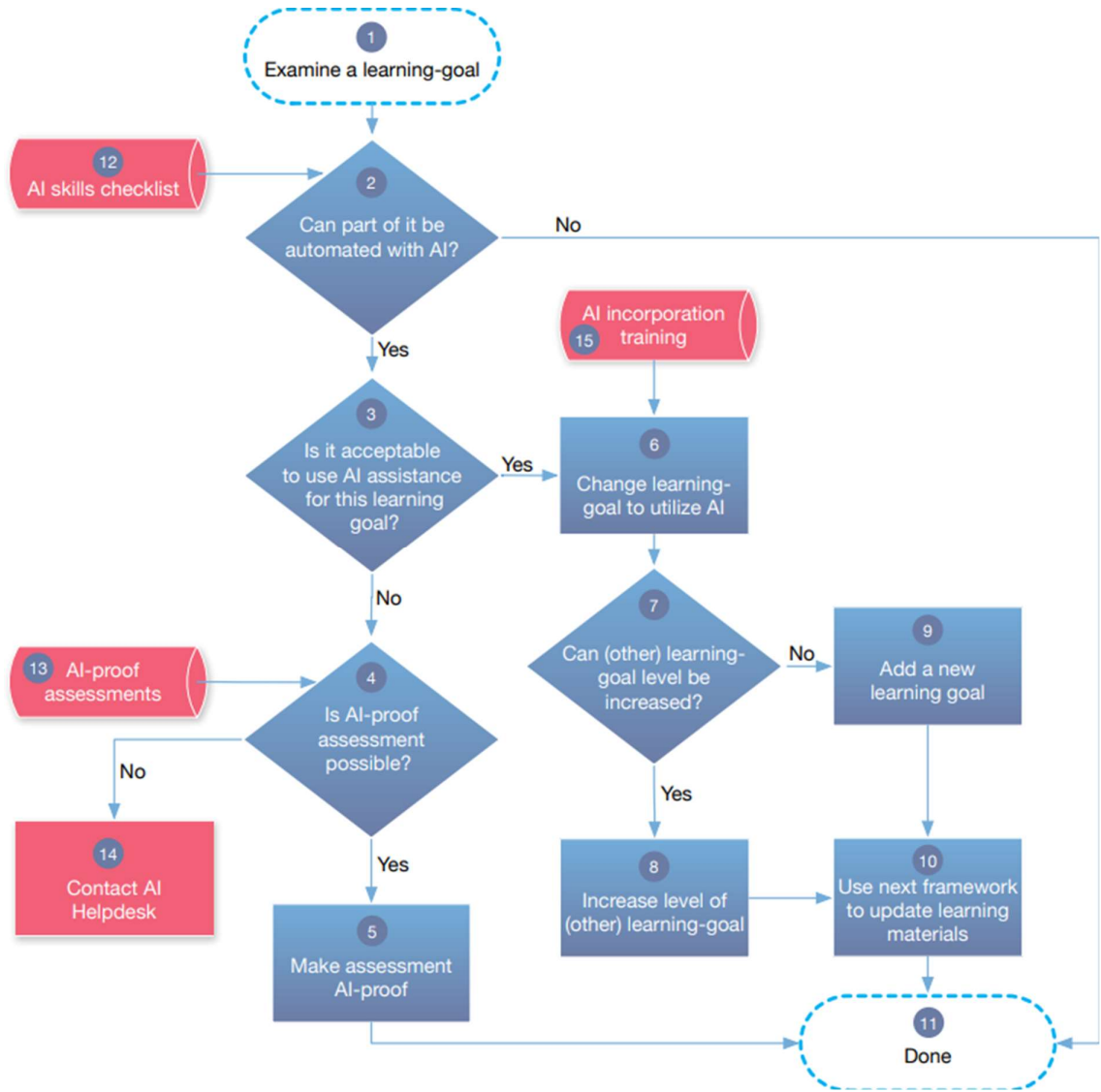

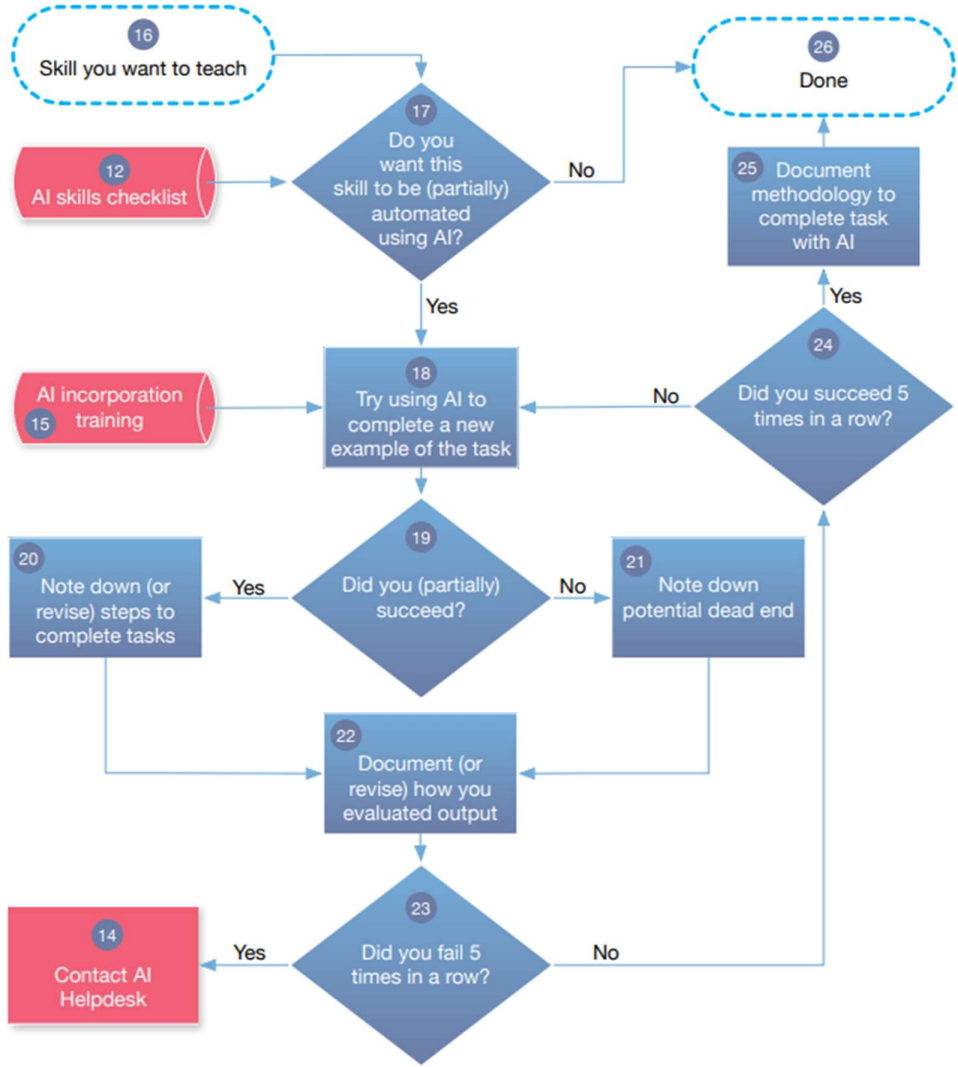

*Figure 2: SELAR – Workflow Learning Materials*